\documentclass[useAMS,usenatbib]{mn2e}
\usepackage{graphicx}
\usepackage[usenames]{color}

\title[A planet transiting a rapidly-rotating A5 star]{Line-profile tomography of exoplanet transits -- II. A gas-giant planet transiting a rapidly-rotating A5 star
\thanks{(Based on observations at  Tautenburg Observatory, McDonald Observatory and the Nordic Optical Telescope)}}
\author[A. Collier Cameron et al]
{
A. Collier Cameron$^{1}$\thanks{E-mail:acc4@st-and.ac.uk},
E. Guenther$^{2}$,
B. Smalley$^{3}$,
I. McDonald$^{3}$,
L. Hebb$^{4}$,
\newauthor
J. Andersen$^{5,6}$,
Th. Augusteijn$^{6}$,
S. C. C. Barros$^{7}$,
D. J. A. Brown$^{1}$,
\newauthor
W. D. Cochran$^{8}$,
M. Endl$^{8}$,
S. J. Fossey$^{9}$,
M. Hartmann$^{2}$,
P. F. L. Maxted$^{3}$,
\newauthor
D. Pollacco$^{7}$,
I. Skillen$^{10}$,
J. Telting$^{6}$,
I. P. Waldmann$^{9}$
and 
R. G. West$^{11}$
\\
$^{1}$SUPA, School of Physics and Astronomy, 
University of St Andrews, 
North Haugh, St Andrews, 
Fife KY16 9SS, UK.\\
$^{2}$Th{\" u}ringer Landessternwarte Tautenburg,
D-07778 Tautenburg,
Germany.\\
$^{3}$Astrophysics Group, Keele University, Staffordshire, ST5 5BG, UK.\\
$^{4}$Vanderbilt University, Department of Physics and Astronomy, Nashville, TN 37235, USA.\\
$^{5}$The Niels Bohr Institute, Astronomy Group, Juliane Maries Vej 30, DK-2100 Copenhagen, Denmark.\\
$^{6}$Nordic Optical Telescope Scientific Association, Apartado 474, ES-38700 Santa Cruz de La Palma, Spain.\\
$^{7}$Astrophysics Research Centre, School of Mathematics \&\ Physics,
QueenÕs University, University Road, Belfast, BT7 1NN, UK.\\
$^{8}$McDonald Observatory, University of Texas, Austin, TX 78712, USA.\\
$^{9}$Department of Physics and Astronomy, University College London, Gower Street, London WC1E 6BT, UK.\\
$^{10}$Isaac Newton Group of Telescopes, Apartado de Correos 321, E-38700 Santa Cruz de la Palma, Tenerife, Spain. \\
$^{11}$Department of Physics and Astronomy, University of Leicester, Leicester, LE1 7RH, UK.\\
}
\begin{document}

\date{Accepted 0000 December 00. Received 0000 December 00; in original form 0000 October 00}

\pagerange{\pageref{firstpage}--\pageref{lastpage}} \pubyear{2008}

\maketitle

\label{firstpage}

\begin{abstract}
Most of our knowledge of extrasolar planets rests on precise radial-velocity measurements, either for direct detection or for confirmation of the planetary origin of photometric transit signals. This has limited our exploration of the parameter space of exoplanet hosts to solar- and later-type, sharp-lined stars. Here we extend the realm of stars with known planetary companions to include hot, fast-rotating stars. Planet-like transits have previously been reported in the lightcurve obtained by the SuperWASP survey of the A5 star HD15082 (WASP-33; $V=8.3$, $v\sin i = 86$ km s$^{-1}$). Here we report further photometry and time-series spectroscopy through three separate transits, which we use to confirm the existence of a gas giant planet with an orbital period of 1.22d in orbit around HD15082. From the photometry and the properties of the planet signal travelling through the spectral line profiles during the transit we directly derive the size of the planet, the inclination and obliquity of its orbital plane, and its retrograde orbital motion relative to the spin of the star. This kind of analysis opens the way to studying the formation of planets around a whole new class of young, early-type stars, hence under different physical conditions and generally in an earlier stage of formation than in sharp-lined late-type stars. The reflex orbital motion of the star caused by the transiting planet is small,  yielding an upper mass limit of 4.1 M$_{\rm Jupiter}$ on the planet. We also find evidence of a third body of sub-stellar mass in the system, which may explain the unusual orbit of the transiting planet. In HD 15082, the stellar line profiles also show evidence of non-radial pulsations, clearly distinct from the planetary transit signal. This raises the intriguing possibility that tides raised by the close-in planet may excite or amplify the pulsations in such stars.

\end{abstract}


\section{Introduction}
\label{sec:intro}

More than 400 extra-solar planets have been discovered orbiting solar- and later-type stars, but very little is known so far about planets orbiting intermediate-mass main-sequence stars. Recent models of planet formation (e.g. \citealt{ida2005}; \citealt{kennedy2008snowline}) suggest that gas-giant formation may peak among the late B stars. Radial-velocity searches for planets orbiting stars in this mass range become feasible only once the star has evolved into a late-type giant (\citealt{hatzes2005giant}; \citealt{johnson2007giants}; \citealt{sato2008giants}), by which time any close-orbiting planets are likely to have been engulfed. Transit signals have been reported for main-sequence A-F-type stars, suggesting the possible presence of close-in planets \citep{christian2006,lister2007,street2007,clarkson2007,kane2008}, but measuring the small reflex motion that would confirm their planetary origin is generally not possible for these line-poor, fast-rotating stars. The hottest star known to passes a transiting planet until now is OGLE2-TR-L9 \citep{snellen2009}, which at spectral type F3 and $v\sin i=33$ km s$^{-1}$ is still not far from solar-type. As a result, we know very little yet about the frequency of giant-planet formation around earlier-type stars, with their harder radiation fields, more distant snow-lines and greater disc masses.

Here we use the established technique of Doppler imaging to confirm the presence of a transiting planet around the bright, fast-rotating star HD 15082 ($V=8.3$; Sp. type A5m, $v \sin i = 86$ km s$^{-1}$). The travelling spectral signature of a planet transiting the disc of the star during the light minimum yields information on the properties of the planet and its orbit.  This complements or supplements that derived from standard photometric analyses, notably its size, retrograde orbit, and non-alignment of the stellar and orbital spin axes.

\section{Discovery of transits and follow-up photometry}
\label{sec:phot}

HD 15082 (WASP-33) is an early-type star in which flat-bottomed, planet-like transits recurring every 1.22 days were discovered by \citet{christian2006} in the WASP survey \citep{pollacco2006}. However, it rotates too rapidly to permit straightforward confirmation of the planet from precise radial-velocity observations, and so was selected for the more sophisticated analysis reported here. 

\begin{figure}
\includegraphics[width=8.8cm]{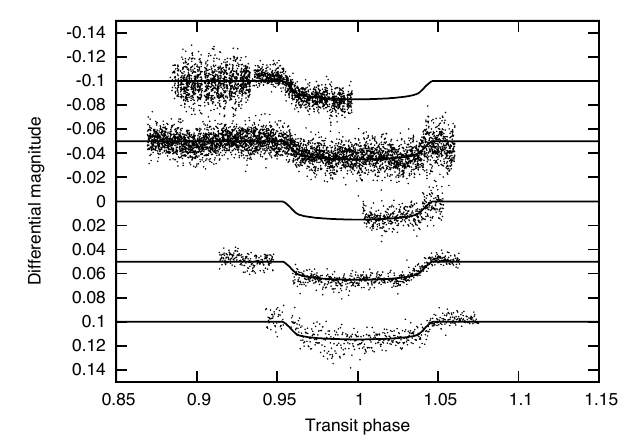}
\caption[]{Transit light curves of HD 15082. Top to bottom: JGT 2006 Nov 13, Keele 2006 Dec 11, Keele 2007 March 20, Mill Hill 2007 Nov 14 (all $R$-band); and Mill Hill 2008 Sep 20 ($I$-band).}
\label{fig:followup_phot}
\end{figure}

Our first step was to follow up the WASP discovery with dedicated photometric observations of HD 15082 to improve the definition of the transit light curve. A partial transit was observed in the $R$ band on 2006 November 13 using the CCD camera of the 0.95-m James Gregory Telescope (JGT) at the St Andrews University Observatory. One complete and one partial $R$-band transit were observed with the 60-cm telescope and CCD camera of the University of Keele \citep{pollacco2007_wasp-3} on 2006 December 11 and 2007 March 20. Finally, complete transits were observed in $R$ and $I$ bands with the 35-cm Schmidt-Cassegrain telescope and CCD camera at the University of London Observatory, Mill Hill \citep{fossey2009hd80606} on the nights of 2007 November 14 and 2008 September 20. In all three instruments the unvignetted field of view includes several bright reference stars, which were used to derive the transit light curves shown in Fig.~\ref{fig:followup_phot} using differential aperture photometry. The total transit duration is 2.72 hours from first to last contact; ingress and egress last 16 minutes, and the transit depth is 0.015 mag in both bands.

\section{Exploratory spectroscopy}
\label{sec:spec1}

\begin{table}
\caption[]{Radial velocities measured with the 2.0-m telescope at TLS.}
\label{tab:rvorbit}
\begin{tabular}{lrrr}
\hline\\
BJD-2450000.0 & Spectrum  &Radial velocity & RV error \\
(days) & SNR & (km s$^{-1}$) & (km s$^{-1}$)\\
\hline\\
4070.4555   &	 62   &       0.5     &       0.8   \\
4070.4736   &	 67   &       2.4     &       0.7   \\
4070.4917   &	 64   &       0.4     &       0.7   \\
4070.5099   &	 50   &       0.6     &       0.6   \\
4070.5298   &	 78   &       1.1     &       0.7   \\
4070.5515   &	 56   &       $-$0.1  &       0.6   \\
4070.5732   &	 65   &       1.5     &       0.7   \\
4071.3650   &	 75   &       0.6     &       0.6   \\
4080.1945   &	 51   &       1.2     &       0.7   \\
4080.2789   &	 51   &       0.3     &       0.7   \\
4080.3785   &	 50   &       $-$0.9  &       0.7   \\
4080.4508   &	 51   &       1.4     &       0.7   \\
4080.5290   &	 48   &       $-$0.9  &       0.7   \\
4082.2761   &	 70   &       0.2     &       0.6   \\
4082.4559   &	 78   &       0.9     &       1.1   \\
4136.2739   &	 72   &       $-$1.2  &       0.8   \\
4136.3863   &	 57   &       $-$0.1  &       0.8   \\
4138.2696   &	 21   &       $-$2.5  &       1.2   \\
4138.2844   &	 20   &       $-$1.7  &       1.5   \\
4156.3454   &	 70   &       $-$1.6  &       0.8   \\
4157.2789   &	 90   &       $-$0.9  &       0.6   \\
4157.3393   &	 81   &       $-$0.7  &       0.8   \\
4161.3098   &	 73   &       $-$0.8  &       0.6   \\
4161.3248   &	 47   &       $-$1.5  &       0.7   \\
4161.3399   &	 31   &       $-$2.3  &       1.3   \\
4161.3547   &	 46   &       $-$1.9  &       1.3   \\
4162.2734   &	 57   &       1.9     &       0.9   \\
4164.2729   &	 76   &       $-$0.9  &       0.7   \\
4165.3654   &	 46   &       $-$1.2  &       0.7   \\
\hline
\end{tabular}
\end{table}

Preliminary radial-velocity (RV) measurements of HD 15082 were obtained at the Th{\" u}ringer Landessternwarte Tautenburg (TLS) in order to determine the mass of the planet from the host star's orbital reflex motion. Observations were made during the periods 2006 November 30 -- December 12 and 2007 February 04 -- 2007 March 05, using the coud\'e\'echelle spectrograph of the 2-m Alfred Jensch telescope at a resolving power of $\lambda/\Delta\lambda=67,000$, 
covering the wavelength range from
4700 to 7400 \AA.

WASP-33 is part of the Tautenburg search for extrasolar
planets (e.g. \citealt{guenther2009}). Radial-velocity measurements are obtained using an iodine cell. As described in detail by \citet{hatzes2005giant}, we fit an initial global wavelength solution to all orders of the spectrum of a ThAr lamp,  then use the lines from the iodine cell to determine the instrumental profile (IP) and the instrumental shift simultaneously with the observations. An accuracy of 1.2 to 1.7 m~s$^{-1}$ is achieved for bright slowly-rotating stars \citep{hatzes2007}. 

\begin{figure}
\includegraphics[width=8.8cm]{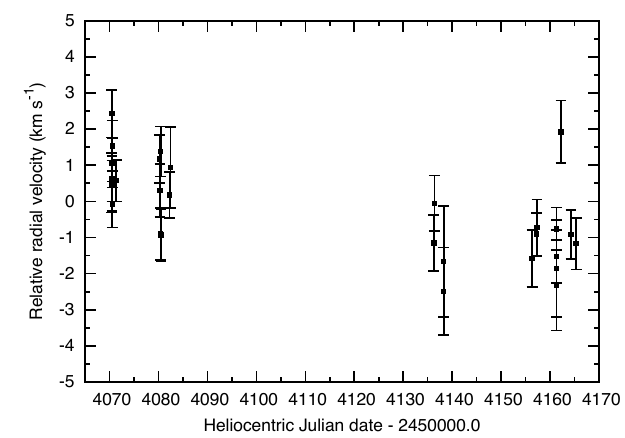}
\includegraphics[width=8.8cm]{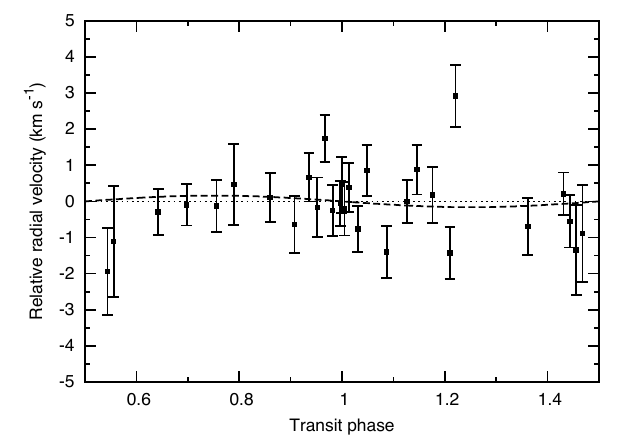}
\caption[]{Radial-velocity curve of HD 15082 obtained with the iodine cell at Tautenburg. The upper panel shows the long-term radial acceleration. The lower panel shows the phased radial velocities with a linear trend of  0.018 km s$^{-1}$ d$^{-1}$ subtracted. The dashed curve represents the best-fitting circular-orbit solution. The zero-point of the radial-velocity scale is arbitrary.}
\label{fig:RVcurve}
\end{figure}

In the case of WASP-33, however, the reduced stellar spectra show extreme rotational broadening, with $v\sin i \simeq 90 \pm 10$ km s$^{-1}$. 
This precludes the direct determination of precise radial velocities relative to the iodine spectrum. Instead, relative radial velocities were computed by cross-correlation with the template spectrum of the star outside the iodine spectral region. 

The results are listed together with the signal-to-noise ratios (SNR) of the individual spectra in Table~\ref{tab:rvorbit}, and the radial-velocity curve is plotted in Fig.~\ref{fig:RVcurve}. There is a long-term trend in the radial velocities over the 95-day span of these observations, corresponding to a radial acceleration of $-18\pm 3$~m~s$^{-1}$~d$^{-1}$ in the centre of mass of the transiting pair. Although most of the stars in the Tautenburg radial-velocity surveys do not exhibit trends at this level \citep{guenther2007}, we are cautious about attributing this trend to another body in the system, because of the star's non-radial pulsations.

The precision of the RV measurements is insufficient for a significant detection of the star's reflex motion on the 1.22-day orbit of the transiting $b$ component from these data alone. 
The RMS scatter about the best-fitting solution is 0.99 km~s$^{-1}$, whereas the mean formally-estimated uncertainty on an individual RV measurement is 0.8 km~s$^{-1}$. The uncertainty in the measured radial-velocity amplitude yields a 99.9 percent upper limit of 4.1 M$_{Jup}$ for the mass of companion $b$, so if the transits are genuinely caused by a body orbiting the host star, it has to be of planetary rather than stellar mass. However, the possibility remained that the transits could be caused by a faint, spatially-unresolved eclipsing-binary companion.

\section{Stellar properties of HD 15082}

\citet{grenier99} classify HD 15082 as an A5mA8F4 star, i.e. A5 from the Ca {\sc ii} $K$ line, A8 from the H lines, and F4 from the metal lines. This would suggest that HD15082 is a classical Am star with overabundances of iron-group metals and underabundances of Ca and Sc \citep{wolff83}.
Stromgren-Crawford $uvby\beta$ photometry \citep{hauck97} yields $T_{\rm eff} = 7430$ K, $\log g = 4.21$ and $[M/H] = 0.21$ via the calibration of
\citet{smalley93}, while Geneva photometry from the General Catalogue of
Photometric Data\footnote {GCPD, http://obswww.unige.ch/gcpd/gcpd.html} yields $T_{eff}=7471\pm 63$~K, $\log g = 4.35 \pm 0.07$ and $[M/H] = 0.08\pm 0.09$ with the \citet{kunzli97cal} calibration.

The template spectrum (taken without the iodine cell) was analysed using the {\sc uclsyn} spectral synthesis package (\citealt{smith92uclsyn}; \citealt{smalley2001uclsyn}). A reasonable fit to the spectrum is obtained with $T_{eff} = 7400 \pm 200$~K, $\log g = 4.3 \pm 0.2$ and $[M/H] = 0.1 \pm 0.2$, in good agreement with the values derived from photometry. No obvious Am characteristics are visible in this spectrum other than slightly weak Ca{\sc ii} $H$\&$K$ lines; a definitive analysis of this issue will require a spectrum with much higher signal-to-noise. The stellar properties are summarized in Table~\ref{wasp33-params}.

\begin{table}
\caption{Stellar parameters of WASP-33 from Tautenburg template spectrum and catalogue data.}
\label{wasp33-params}
\begin{tabular}{cc} \hline
Parameter     & Value \\ \hline
RA (2000.0)   & 02m26m51.06s \\
Dec (2000.0)  & +37{\degr}33{\arcmin}01.7{\arcsec} \\
$V$~{mag.}    & 8.3 \\
Spectral Type & kA5hA8mF4 \\
Distance      & 116 $\pm$ 16 pc \\
$T_{\rm eff}$ & 7430 $\pm$ 100 K \\
$\log g$      & 4.3 $\pm$ 0.2 \\
$v \sin i$    &  90 $\pm$ 10 km s$^{-1}$ \\
{[M/H]}       &  0.1 $\pm$ 0.2 \\
\hline
\end{tabular}

Notes:
Spectral type from \citet{grenier99}\\
Distance from Hipparcos \citep{perryman97}
\end{table}

\section{Time-resolved transit spectra}
\label{sec:transpec}

During a transit, a planet blocks a small part of the stellar photosphere. The missing starlight no longer contributes to the overall spectral line profile, but appears as a resolved ``bump" within the rotational broadening function, in the same way as dark star spots distort the line profiles of a rotating star \citep{vogt83doppler}. If such a signature is present and matches the transit geometry deduced from photometry, the presence of a planet may be confirmed even if the host star's reflex orbit cannot be reliably measured. 
\citet{jenkins2010} have also noted recently that the Rossiter effect provides a powerful method for distinguishing planets from blends in cases where classical methods of confirmation cannot be used.

The spectral signature of a blended stellar binary would look very different.  To produce a transit with a fractional depth of 0.01, even an object with a very deep eclipse should contribute at least one percent of the total system light. At a period of 1.2 days, the radial-velocity amplitude of the primary star will be of order 100  km s$^{-1}$, giving a velocity drift of order tens of km/sec within the 3-hour duration of a typical transit. When the light of the brighter foreground  star is subtracted, such a background signature is easily detectable in the residual line profiles. The Rossiter distortion will be imposed on the rotation profile of this drifting stellar signature, not on that of the foreground star. 

\begin{figure}
\includegraphics[width=9cm]{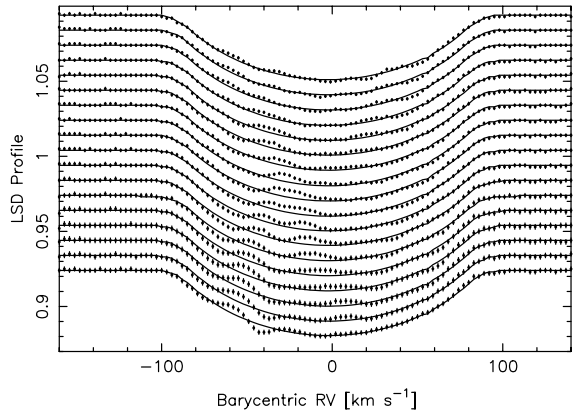}
\caption[]{Least-squares deconvolved profiles of HD 15082 during the planetary transit of 2009 December 08 observed at the NOT. Data are shown as points with error bars. The limb-darkened stellar rotation profile is shown as a solid line. Time increases vertically. The planet signature is the positive bump that migrates from $v\simeq 5$ km~s$^{-1}$ in the first (bottom) profile to $v\simeq -40$ km~s$^{-1}$ in the last (top) profile shown.}
\label{fig:profiles}
\end{figure}

\begin{figure*}
\includegraphics[width=8.8cm]{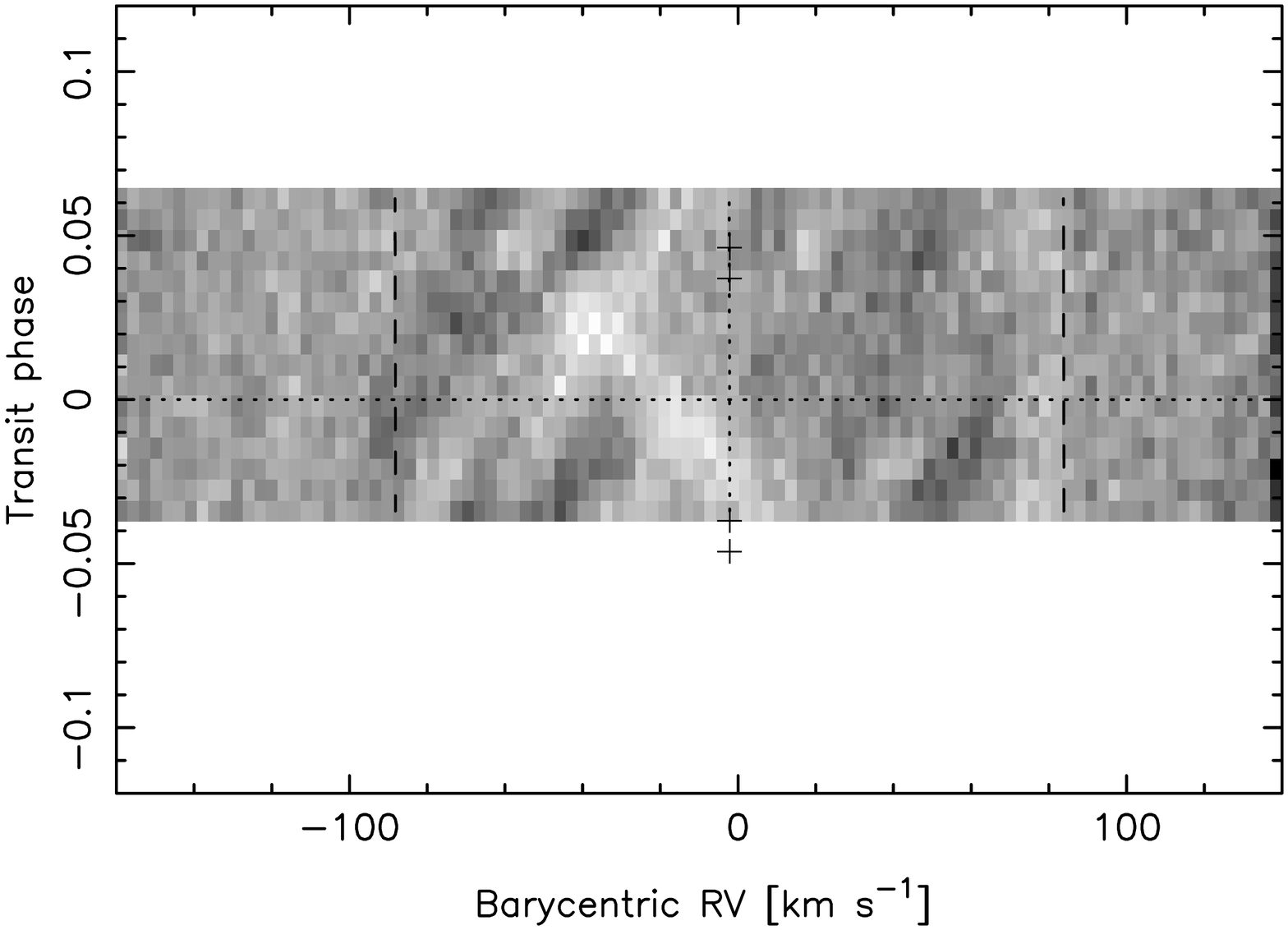}
\includegraphics[width=8.8cm]{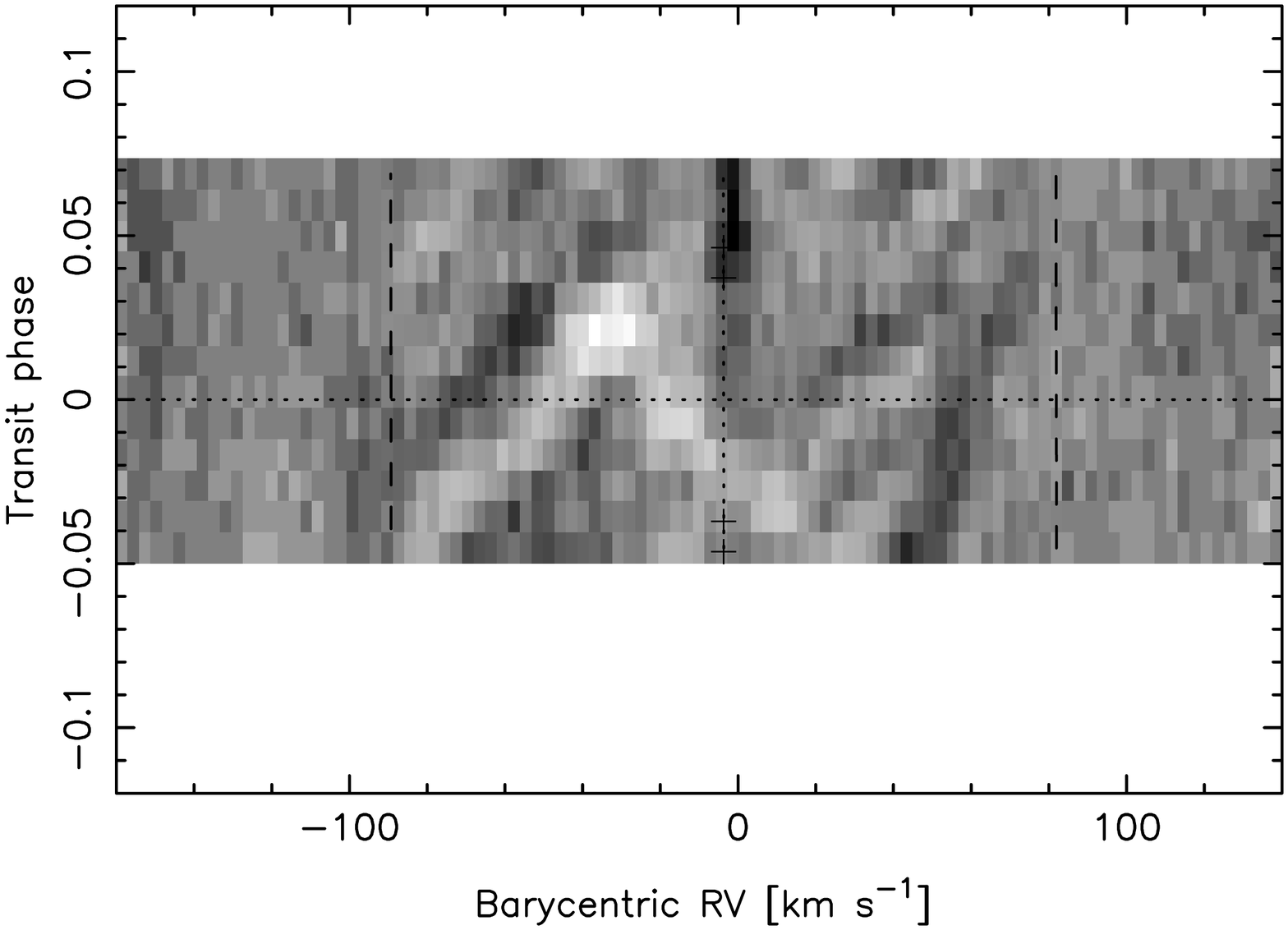}
\includegraphics[width=8.8cm]{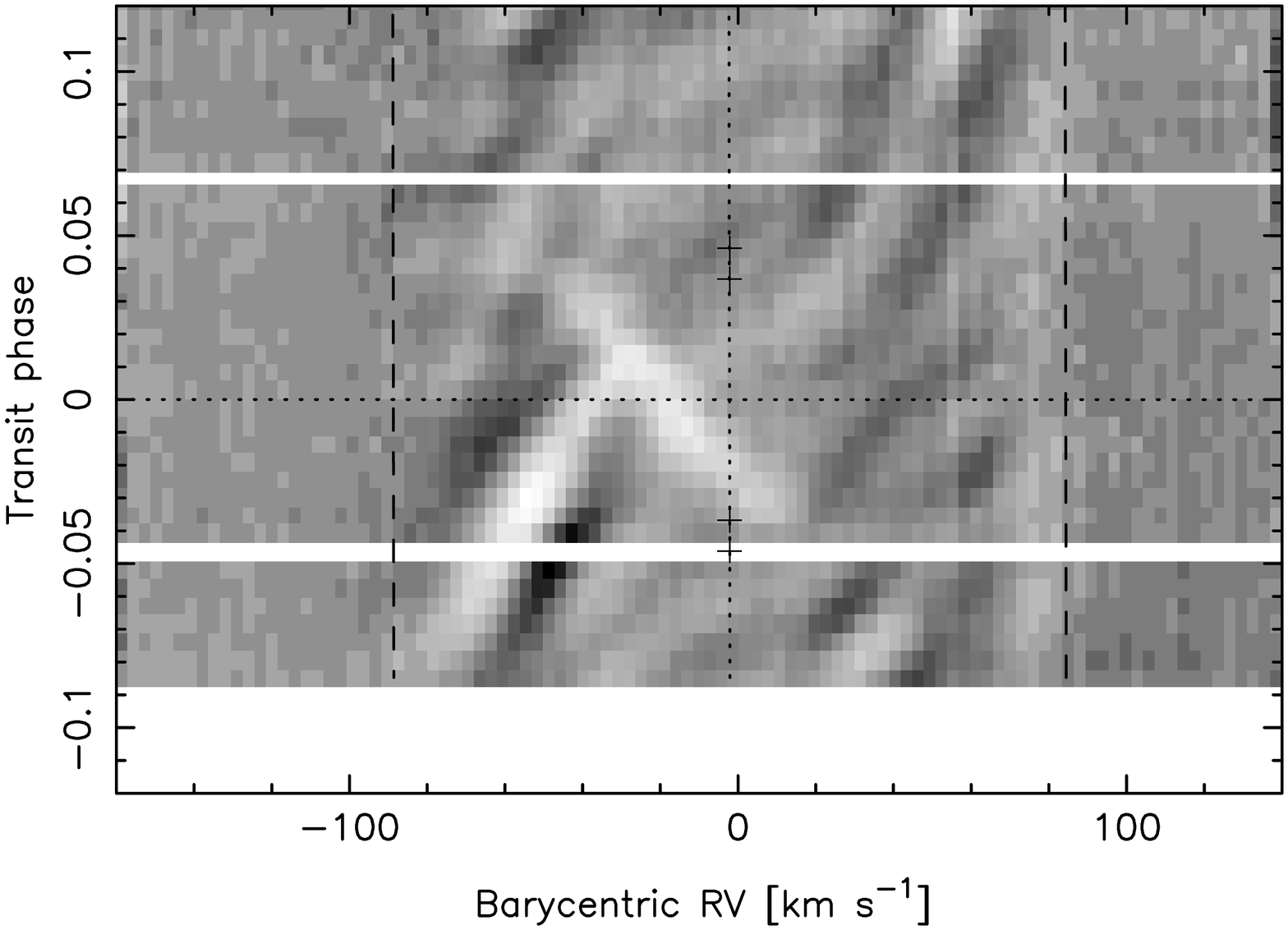}
\includegraphics[width=8.8cm]{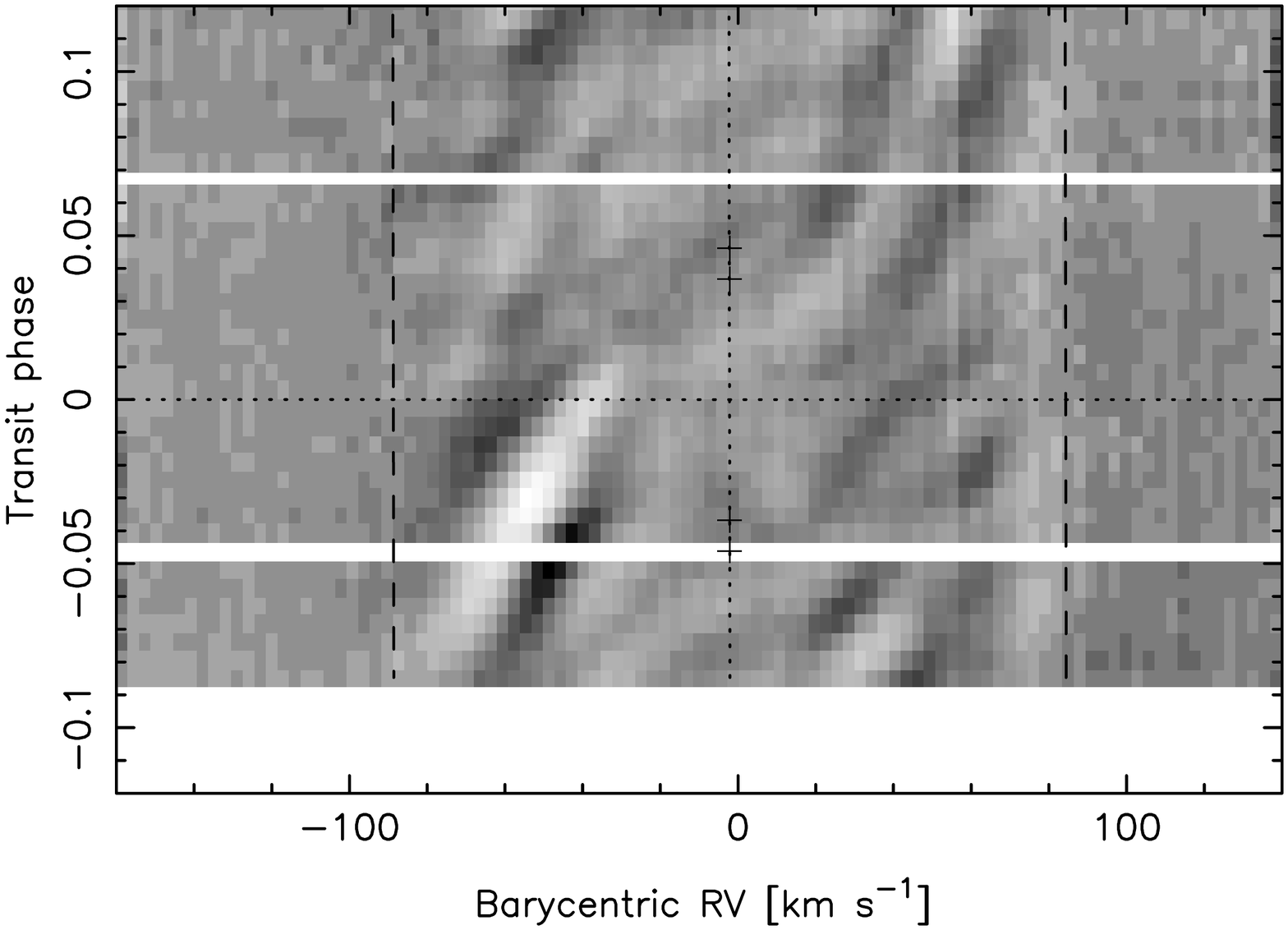}
\caption[]{Time series of the residual average spectral line profile of HD 15082 during the transits of 2008 October 18 at Tautenburg (upper left); 2008 November 11/12 at McDonald (upper right); and 2009 December 8 at the NOT (lower left). Wavelength or radial velocity increases from left to right, time from bottom to top. The limb-darkened rotation profile of the star itself has been subtracted between the dashed lines at $\pm v\sin i$, and the velocity of the star is denoted by a dotted vertical line with the four contact times of the transit indicated by "+" symbols. The planetary signal is
the bright feature that moves from right-to-left coincident with the duration
of the photometric transit. In the lower-right panel, the model planet signature has been subtracted from the NOT data, leaving only the pattern of non-radial pulsations. 
}
\label{fig:trailed}
\end{figure*}

A new set of \'echelle spectra was therefore obtained at TLS on the night of 2008 October 18. Sixteen spectra with 600s integration time were obtained at $R=40,000$ from 4708 to 7004 \AA without the iodine cell, starting just before ingress and ending after egress. A second dataset was obtained with the CS23 \'echelle spectrograph on the 2.7-m Harlan J. Smith Telescope at McDonald Observatory during the night of 2008 November 11/12, with a spectral resolving power of $R=60,000$ from 3837 to 10209 \AA\ and exposure times of 900s. This series also covers the entire transit. A third sequence with 540-s time resolution was obtained with the FIES spectrograph at the Nordic Optical Telescope (NOT) on La Palma on 2009 December 8, extending for $\sim2$ hours before and after the transit to characterise better the stellar pulsations seen earlier and discussed below.  The wavelength range was 3911 to 7271 \AA\ with $R=67000$.

For each dataset, we combined the information in all the spectral lines in each echellogram into a single, composite profile with a high signal-to-noise ratio using Least-Squares Deconvolution (LSD). This method, pioneered by \citet{donati97zdi}, computes the line profile which, when convolved with a forest of appropriately-weighted delta functions at the wavelengths of known spectral lines, gives an optimal inverse variance-weighted fit to the data. The velocity scale is tied to the  reference frame of the Earth's atmosphere along the observer's line of sight using a separate LSD calculation based on a mask of narrow telluric absorption lines. This compensates for any drift in the instrumental profile to an accuracy of order 10 to 20 m~s$^{-1}$. This is more than sufficient for tracking features in the line profile that move through several tens of km~s$^{-1}$ during the transit. The signal-to-noise ratios of the composite profiles are 1070 for the TLS data, 1730 for the McDonald data, and 1390 for the NOT data. The resulting time-series spectra during the NOT transit are shown in one-dimensional form Figure~\ref{fig:profiles}. The trailed spectra for all three data sets are shown in grey-scale form in Figure~\ref{fig:trailed}, with the mean line profile of the star itself subtracted. They exhibit a clear Doppler signature of a planet moving in the retrograde direction across the stellar rotation profile. They also show the characteristic pattern of stellar non-radial pulsations carried in the prograde direction by the stellar rotation. We discuss the planetary signal further in the following and the pulsations in Section~\ref{sec:pulsation} below.

\begin{table*}
\caption[]{System parameters of HD 15082 as derived from our photometry, radial velocities and transit spectroscopy.}
\label{tab:params}
\begin{tabular}{lcccc}
\hline\\
Parameter & Phot+RV & TLS & McD & NOT \\
\hline\\
Epoch of mid-transit, $T_c$ (HJD)  & $2454163.22373 \pm  0.00026$ & - & - & - \\
Orbital period, $P$ (days) & $1.2198669 \pm  0.0000012$ & - & - & - \\

Planet/star radius ratio, $R_p/R_s$ & $0.1066\pm  0.0009$ & $0.1072\pm  0.0006$ & $0.1071\pm  0.0006$ & $0.1073\pm  0.0006$ \\
Scaled stellar radius, $R_s/a$ & $0.2640\pm  0.0057$ & $0.2673\pm  0.0015$ & $0.2653\pm  0.0016$ & $0.2661\pm 0.0015$ \\
Impact parameter $b$ & 0.155$^{+ 0.100}_{-0.120}$ & $0.218\pm  0.008$ & $0.176\pm  0.010$ & $0.203\pm  0.007$   \\
Centre-of-mass velocity, $\gamma$ (km s$^{-1}$) &  & $-2.19\pm 0.09$ & $-3.69\pm 0.09$ & $-2.11\pm  0.05$ \\
Radial acceleration, $d\gamma/dt$ (km s$^{-1}$ d$^{-1}$)& $-0.0183\pm 0.0035$& - & - & - \\
Radial-velocity amplitude, $K_s$ (km s$^{-1}$) & $ < 0.59 $ & - & - & - \\
Stellar rotation speed, $v\sin i$ (km s$^{-1}$) & - & $86.04\pm 0.13$ & $85.64\pm  0.13$ & $86.48\pm  0.06$ \\
Projected obliquity, $\lambda$ (degrees) & - & $251.2\pm 1.0$ & $254.2\pm 1.2$ & $251.6\pm 0.7$ \\
Intrinsic linewidth, $v_{\rm fwhm}$ (km s$^{-1}$) & - & $16.2\pm 0.5$ & $19.2\pm 0.6$ & $18.1\pm 0.3$ \\

Stellar density, $\rho_s/\rho_\odot$ & $0.497\pm  0.024$ & $0.472\pm  0.008$ & $0.483\pm  0.008$ & $0.479\pm  0.007$ \\
Stellar mass, $M_s/M_\odot$& $1.495\pm  0.031$ & $1.497\pm  0.029$ & $1.495\pm  0.029$ & $1.500\pm  0.030$ \\
Stellar radius, $R_s/R_\odot$& $1.444\pm  0.034$ & $1.469\pm  0.013$ & $1.458\pm  0.013$ & $1.464\pm  0.013$ \\
Planet mass, $M_p/M_{\rm Jup}$ & $ < 4.1 $ & - & - & - \\
Planet radius, $R_p/R_{\rm Jup}$ & $1.497\pm  0.045$ & $1.533\pm  0.016$ &$1.519\pm  0.016$ & $1.527\pm  0.015$ \\

Orbital separation, $a$ (AU) & $0.02555\pm  0.00017$ & $0.02556\pm  0.00017$ &  $0.02555\pm  0.00017$ &  $0.02557\pm  0.00014$\\
Orbital inclination, $i$ (degrees) & $87.67\pm 1.81$  & $86.67\pm  0.13$ & $87.33\pm  0.15$ & $86.91\pm  0.11$ \\

\hline
\end{tabular}
\end{table*}

Fig.~\ref{fig:trailed} shows the wealth of direct information on the planet's motion that is available from these data: First, ingress (bottom right) clearly occurs over the receding limb of the star, egress (top left) over the approaching limb -- immediate evidence that the orbital motion of the planet is opposite to the rotation of the star, i.e. retrograde. The velocity width of the planet signal relative to that of the stellar line profile corresponds to the length of the chord traversed by the planet relative to the diameter of the stellar disc, i.e. the impact parameter of the transit, related to the orbital inclination. Similarly, ingress clearly occurs closer to the spin axis of the star than the egress, proof that the orbit is inclined relative to the stellar equator as well as to the line of sight, and that the trace is not due to a spot on the surface of the star (the transit speed confirms this as well). Finally, the width and depth of the planet trace are direct indicators of the radius and surface area of the planet, when allowance is made for the effects of the PSF, natural line width, and acceleration of the planet during each exposure.

Information on the rotation of the star and the direction of its spin axis relative to that of the orbit can also be derived from accurate radial-velocity data through the Rossiter-McLaughlin effect (e.g. \citealt{anderson2010wasp17}), but is available in a far more direct and less model-dependent way from the line-profile analysis discussed here, when the stellar rotation is large enough. The radius of the star and the inclination of the orbit can be determined both by the standard analysis of the transit light curve and by the Doppler imaging performed here, which allows at least a consistency check. Which of the methods yields the more accurate results depends on the system in question and the quality of the two datasets.

\section{Combined analysis}
\label{sec:combined}

In this section we integrate all the available data into a combined analysis to determine the properties of the planet and the star. This includes the WASP data from 2004 discussed by \citep{christian2006} plus a further 3658 observations obtained by the WASP survey in 2007.

For the light curves, we modelled the transit geometry and orbital parameters with the Markov-chain Monte-Carlo parameter fitting code described in detail by \citet{cameron2007mcmc} and \citet{pollacco2007_wasp-3}. Five parameters are determined: the epoch $T_0$ of mid-transit, the orbital period $P$, the planet/star radius ratio $R_p/R_s$, the scaled stellar radius $R_s/a$ (where $a$ is the orbital semi-major axis) and the impact parameter $b=a \cos i/R_s$, where $i$ is the orbital inclination. Model light curves are computed using the analytic formulation of \citet{mandel2002} in the small-planet approximation with non-linear limb darkening. Slow small-scale photometric trends are removed in the process. 

The stellar density $\rho$ is derived from $R_s/a$ and $P$ via Kepler's third Law \citep{seager2003}:
\begin{equation}
\rho_s=\frac{3 M_s}{4\pi R_s^3}=\frac{3\pi}{G P^2}\left(\frac{a}{R_s}\right)^3.
\end{equation}
The stellar mass is estimated at each step from the empirical calibration of \citet{torres2009} as modified by \citet{enoch2009}. The radius of the star, and hence that of the planet, then follow. 

A circular orbit with a constant radial-acceleration term was fit to the radial-velocity data to derive the stellar reflex velocity amplitude $K_s$. At each step in the Markov chain, the centre-of-mass velocity $\gamma$ at the fiducial epoch of transit is computed as the mean of the residuals. 

The MCMC solution combining the photometry and the Tautenburg radial velocities is given in the first column of Table~\ref{tab:params}.
The secular trend in the values of $\gamma$ from 2006 and 2007 is found to be $-18\pm 3$ m~s$^{-1}$~d$^{-1}$. If attributed to a binary companion (denoted as component $c$), the derived mass would be of order $M_c = 0.036 (a_c/AU)^{2}$ M$_\odot$. The outer period appears to be at least 200 days, suggesting that component $c$, if real, should have a mass at least 20 times that of Jupiter. This interpretation should, however, be treated with caution. The non-radial pulsations distort the line-profile shapes strongly, introducing substantial radial-velocity shifts which could introduce a spurious radial acceleration into the orbital solution. The introduction of a linear trend into the model reduces the RMS scatter about the fitted solution from 1.30 km~s$^{-1}$ to 0.99 km~s$^{-1}$. An $F$-test on the residuals about the fitted solution with and without the linear trend returns a 17 percent probability that their variances are not significantly different.

For the line profile analysis we incorporated each of the three time-resolved spectral sequences in the MCMC solution. We used the profile decomposition method of \citet{cameron2010rm} to determine the rotation speed of the star and the projected orientation of the orbital and stellar spin axes. The model line profile is a limb-darkened rotation profile convolved with a Gaussian corresponding to the intrinsic photospheric line profile plus instrumental broadening. The planet's Doppler shadow is a travelling Gaussian of the same width, computed as described by \citet{mandel2002}. The changing form of the line profile is thus modelled with five additional MCMC parameters \citep{cameron2010rm}: the stellar rotation speed $v\sin i$, the difference $\lambda$ between the position angles of the orbital and stellar rotation axes in the plane of the sky; the impact parameter $b$ of the planet's trajectory; the width $v_{\rm FWHM}$ of the Gaussian intrinsic profile; and a nightly zero-point correction $\gamma$ to the centre-of-mass velocity at the time of observation. The velocity difference between the planet shadow at ingress and egress relative to the stellar velocity provides a better constraint on the value of the impact parameter $b$ than the photometry alone, as noted above: $\Delta v_{\rm i,e}=b\sin\lambda\pm\sqrt{1-b^2}\cos\lambda$.

Individual solutions are given for each set of transit spectroscopy  in Table~\ref{tab:params}, combined with the photometry and Tautenburg radial velocities. The system centre-of-mass velocity differs slightly when different sets of transit spectra are used. This is partly attributable to the outer orbit, and possibly to zero-point differences between the different instruments. The non-radial pulsations, however, are likely to be the dominant additional source of error in the profile fits. They introduce line-profile distortions which cannot be accounted for in the modelling procedure, and whose  amplitudes approach that of the planet's Doppler shadow. The parameters $b$, $v\sin i$, $\lambda$ and $v_{\rm FWHM}$ described in the paragraph above are the ones most directly affected by the pulsations, and indeed their fitted values disagree by several  sigma among the three individual sets of observations. They propagate directly into the derived value for the orbital inclination, which also exhibits a scatter greater than expected  from the formal errors returned by MCMC. For these parameters, the scatter among the different columns is a better indicator of the true uncertainty than the formal errors associated with the individual values. For all other quantities, the mutual agreement is excellent. 

\begin{figure}
\includegraphics[width=8.8cm]{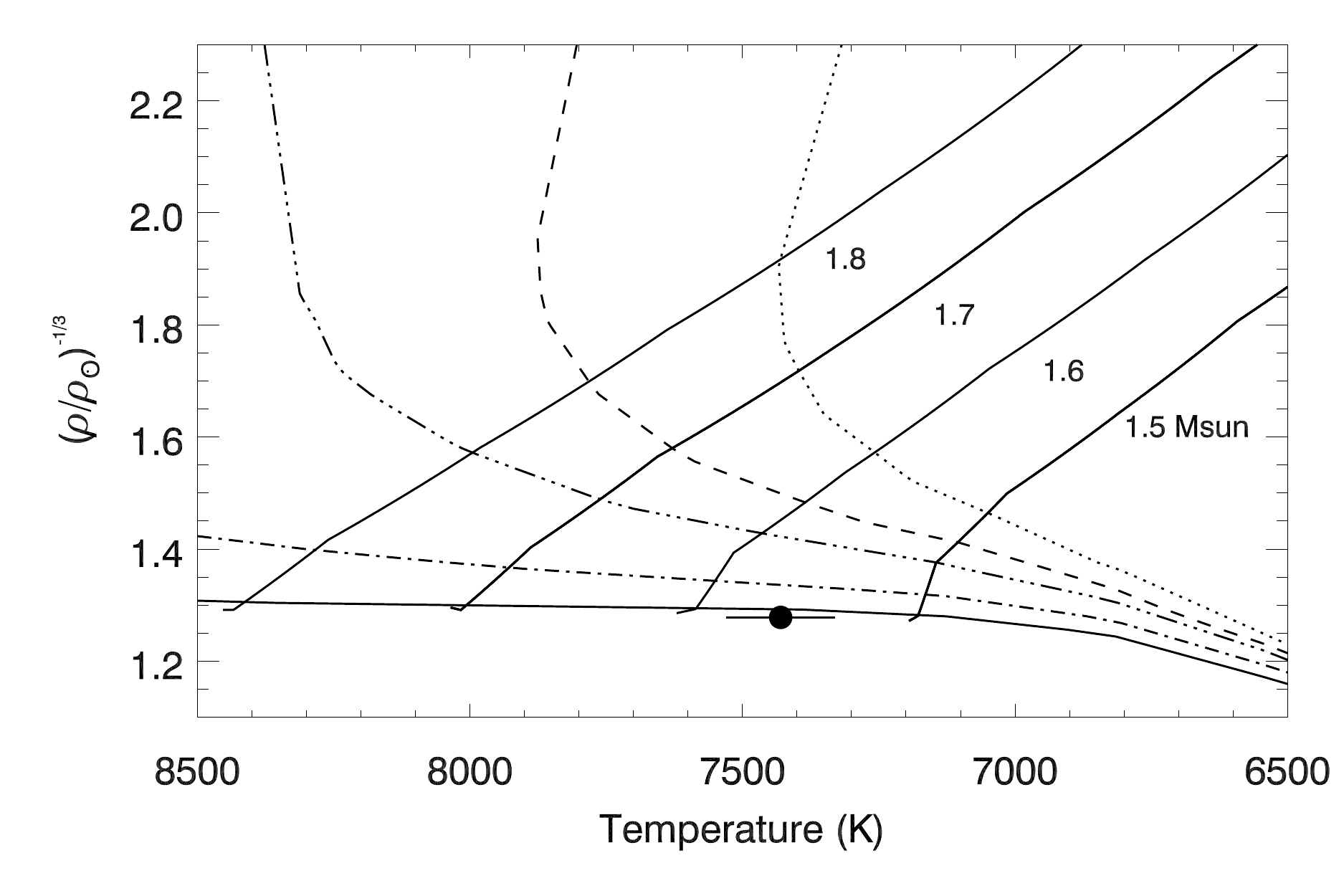}
\caption[]{The effective temperature and inverse cube root of density of HD 15082 are shown in relation to the evolutionary tracks of \citet{girardi2000} for stars of mass 1.5, 1.6, 1.7 and 1.8 M$_\odot$, at ages of 63 Myr (solid), 320 Myr (dot-dash), 630 Myr (triple dot-dash), 795 Myr (dashed) and 1 Gyr (dotted). The error bar on the stellar density is smaller than the plot symbol. A mass of $1.55\pm 0.04$ M$_\odot$ and an age less than 25 Myr are indicated. The tracks and isochrones are computed for [Fe/H]=0.08, corresponding to heavy-element abundance  Z=0.0228. Decreasing [Fe/H] by 0.09 dex lowers the zero-age main-sequence value of $(\rho/\rho_\odot)^{-1/3}$ by 0.03 for the youngest isochrone, and raises $T_{\rm eff}$ by 150K for a 1.5M$_\odot$ stellar model. 
}
\label{fig:wasp33_evol}
\end{figure}

\begin{figure}
\includegraphics[width=8.8cm]{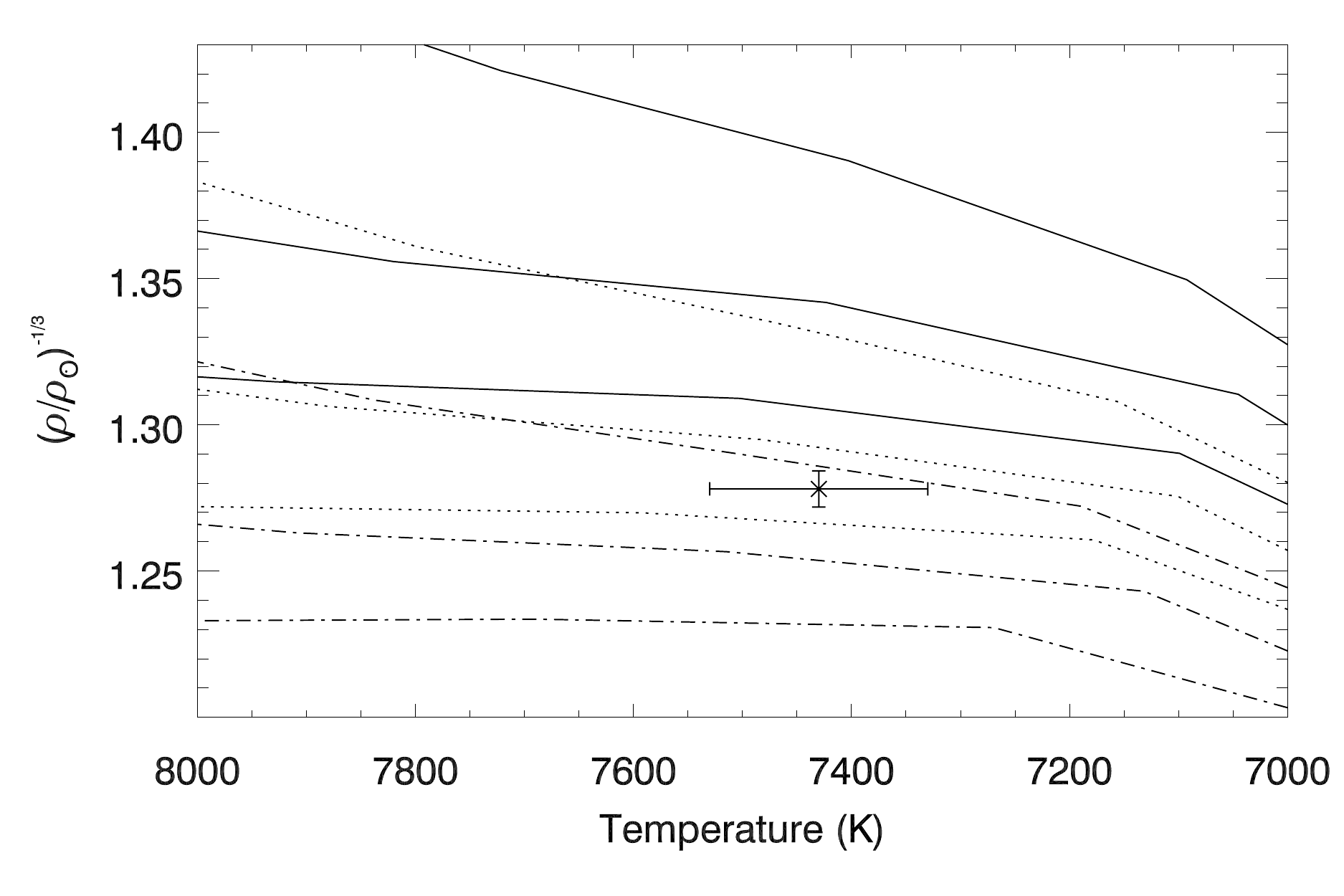}
\caption[]{
The effective temperature and inverse cube root of density of HD 15082 are shown in relation to the Yale-Yonsei isochrones \citep{demarque2004} for stars of [M/H] =  -0.10 (dot-dashed lines), 0.00 (dotted lines) and 0.10 (solid lines). The isochrone ages are 50, 100 and 400 Myr (bottom to top) in each case. The observed stellar density is clearly inconsistent with the upper part of the error range on [M/H] quoted in Table~\ref{wasp33-params}; the tracks shown here match the range from the nominal [M/H]=0.1 to the one-sigma lower limit [M/H]=-0.1. If the star has solar metallicity, it is about 90 Myr old.  If it it 0.1 dex sub-solar, the age increases to ~300 Myr. Given the lower limit on the metallicity derived from the stellar spectra, we infer an upper age limit of $\sim 400$ Myr.
}
\label{fig:wasp33_metalage}
\end{figure}

Even including the uncertainties introduced by the pulsations, the impact parameter is very well determined by the trailed spectra. With the help of the photometry, the stellar density is then determined to a precision of order 2 percent. This allows the stellar age and mass to be estimated in relation to isochrones and evolutionary tracks plotted in the ($T_{\rm eff}, (\rho_*/\rho_\odot)^{-1/3}$) plane \citep{sozzetti2007}. In Fig.~\ref{fig:wasp33_evol} we show the location of HD 15082 in relation to the evolutionary models of \citet{girardi2000} for stars with [Fe/H]=0.08 and masses in the range 1.5 to 1.8 M$_\odot$. The stellar mass is found to be $1.55\pm 0.04$ M$_\odot$, in good agreement with the value
$1.50\pm 0.03$ M$_\odot$
obtained from the empirical calibration built into the MCMC analysis, as listed in Table~\ref{tab:params}.
If we use evolutionary tracks interpolated for  [M/H]=-0.1, we obtain a stellar mass of 1.50 M$_\odot$, which is slightly closer to the value obtained from the eclipsing-binary calibration.

The stellar density is seen to be very close to the main-sequence value for stars in this mass range. Comparison with the Yonsei-Yale isochrones of \citet{demarque2004} shows HD 15082 to be in the early part of its main-sequence life, with an isochrone age no greater than 400 Myr or so after making allowance for a 0.2 dex uncertainty in the metallicity (Fig.~\ref{fig:wasp33_metalage}). It is too dense to lie on or above the main sequence for metallicities greater than solar.

Assuming a stellar $T_{\rm eff}=7430\pm 100$K, zero albedo and uniform redistribution of incident flux, the equilibrium temperature of the planet is $2710\pm 50$ K.

\section{Non-radial pulsations in the host star}
\label{sec:pulsation}

Removing the planet's Doppler shadow from the trailed spectra highlights the underlying pattern of non-radial pulsations of the star. Note that the radial component of the pulsations vanishes near the centre of the line profile (Fig.~\ref{fig:trailed}, lower right panel), a behaviour indicative of g-modes. Overall, the pattern appears to be that of a sectoral non-radial mode with $\ell = 4\pm 2$, suggesting that the star is a $\gamma$ Dor-type variable. The pulsation amplitude appears greater in the blue wing of the profile than the red. This phenomenon was noted earlier by e.g. \citet{uytterhoeven2008puls} who studied HD 49434, a star remarkably similar in $T_{\rm eff}$ and $v\sin i$ to HD 15082. HD 49434 exhibits several pulsation modes with frequencies similar to the $\sim 4$ cycles/day tidal forcing frequency imposed on HD 15082 by its planet. It has been suggested \citep{willems2002} that stellar pulsation modes may be excited by resonant tidal forcing by a close binary companion, and \citet{fekel2003} reported that a surprisingly high fraction of $\gamma$ Dor pulsators have close binary companions. Similar resonances are invoked in the study of dissipative tidal spin-orbit coupling in hot-Jupiter systems (\citealt{ogilvie207tide}; \citealt{barker2008tide}). Whether a retrograde planet  can excite pulsations in a star rotating in the opposite direction remains an open question, however.

\section{Summary and conclusions}

Our time-series spectroscopy has provided the first confirmation of a gas-giant planet transiting a rapidly-rotating A-type main-sequence star, proving that A stars do occasionally harbour close-in planets. The result opens up a new part of the parameter space of exoplanet host stars, where planets are formed under different physical conditions than around solar-type stars, and on shorter time-scales. It will be interesting to compare the properties of planets orbiting young and intermediate-age stars as determined through the Doppler imaging technique to study the early dynamical evolution of planetary orbits in different environments.

With a radius of  1.46~R$_{\rm Jup}$ HD 15082b  is one of the more bloated exoplanets yet discovered. The large radii of some of the  hot Jupiters, notably TrES-4b \citep{mandushev2007tres4}, WASP-12b \citep{hebb2009wasp12} and WASP-17b \citep{anderson2010wasp17}  remain a challenge to theories of planetary structure. Some of the leading contenders are dissipation of tidal energy in the interior of a planet during orbit circularisation \citep{bodenheimer2001radii}, advection of a small fraction of incident stellar flux into the deep interior \citep{guillot2002irrad}, the presence or absence of a substantial  rock/ice core \citep{fortney2007radii} or some combination of the above, with a possible dependence of core mass on host-star metallicity \citep{fressin2007radii}. 

Without a measurable radial-velocity orbit, the mass and density of the planet remain unknown, making it difficult to estimate either its core mass or its orbital eccentricity.
In its 1.2-day orbit, however, HD 15082b is the most strongly-irradiated planet yet found, with an equilibrium temperature of order 2710 K assuming isotropic re-radiation and a low albedo. This may be at least partly responsible for the planet's large radius (e.g. \citet{guillot2002irrad}). At $V=8.3$, HD 15082 is also the fifth-brightest star yet found to host a transiting planet. This will make it an attractive target for future infrared secondary-eclipse studies \citep{charbonneau2005,deming2005,knutson2007}. In particular, the displacement of the secondary eclipse from phase 0.5 and the duration of the secondary eclipse will allow the orbital eccentricity and argument of pericentre to be determined \citep{carter2008}. 

HD 15082b joins the growing ranks of retrograde planets, notably WASP-17b \citep{anderson2010wasp17} and HAT-P-7b (\citealt{winn2009_hat-p-7}; \citealt{narita2009_hat-p-7}). These and other strongly misaligned planets present a challenge to theories of disc migration (e.g. \citealt{lin96}). Since the star and the planets form from the same disc, their respective spin and orbital angular momentum vectors should be aligned. The misaligned planets are, however, plausibly explained by alternative theories involving combinations of planet-planet scattering \citep{rasio96,weidenschilling96}, Kozai oscillations \citep{kozai62,wu2003kozai} and tidal damping \citep{eggleton2001,fabrycky2007kozai,nagasawa2008kozai}. A key future test of these theories will involve searching for additional, exterior bodies in these systems, which might have excited Kozai oscillations in the past. Continued radial-velocity monitoring of this star, and a deep search for faint common-proper-motion companions, would therefore be of interest.

The host star is a non-radial pulsator, apparently of the $\gamma$ Dor type. A full study of the pulsation modes in HD 15082 is beyond the scope of the present paper, but could in the future provide independent confirmation of the stellar density derived here from the transit geometry. 

\section*{Acknowledgments}

This work was based on observations made on the 2-m Alfred Jensch Telescope of the Thu¬ringer Landessternwarte Tautenburg, and on the 2.7-m Harlan J. Smith Telescope at the McDonald Observatory of the University of Texas. It was also based on observations made with the Nordic Optical Telescope, operated on the island of La Palma jointly by Denmark, Finland, Iceland, Norway, and Sweden, in the Spanish Observatorio del Roque de los Muchachos of the Instituto de Astrofisica de Canarias. This research has made use of the ESO Data Archive, the NASA Astrophysics Data System, and the VizieR catalogue access tool, CDS, Strasbourg, France.  SJF and IPW would like to thank Cherry Ng, Dan Smith, Yudish Ramanjooloo and the UCL `ExoCafe' team of undergraduate observers for assistance. We also acknowledge the constructive and insightful comments provided by an anonymous referee.

\bibliographystyle{mn2e}

\begin{thebibliography}{99}

\bibitem[\protect\citeauthoryear{Anderson et 
al.}{2010}]{anderson2010wasp17} Anderson D.~R., et al., 2010, ApJ, 709, 
159 


\bibitem[\protect\citeauthoryear{Barker 
\& Ogilvie}{2009}]{barker2008tide} Barker A.~J., Ogilvie G.~I., 2009, MNRAS, 395, 2268 

\bibitem[\protect\citeauthoryear{Bodenheimer, Lin, 
\& Mardling}{2001}]{bodenheimer2001radii} Bodenheimer P., Lin D.~N.~C., Mardling R.~A., 2001, ApJ, 548, 466 

\bibitem[\protect\citeauthoryear{Carter et al.}{2008}]{carter2008} 
Carter J.~A., Yee J.~C., Eastman J., Gaudi B.~S., Winn J.~N., 2008, ApJ, 
689, 499 


\bibitem[\protect\citeauthoryear{Charbonneau et 
al.}{2005}]{charbonneau2005} Charbonneau D., et al., 2005, ApJ, 626, 
523 


\bibitem[\protect\citeauthoryear{Christian et 
al.}{2006}]{christian2006} Christian D.~J., et al., 2006, MNRAS, 372, 
1117 

\bibitem[\protect\citeauthoryear{Clarkson et 
al.}{2007}]{clarkson2007} Clarkson W.~I., et al., 2007, MNRAS, 381, 
851 


\bibitem[\protect\citeauthoryear{Collier Cameron et 
al.}{2007}]{cameron2007mcmc} Collier Cameron A., et al., 2007, MNRAS, 
380, 1230 

\bibitem[\protect\citeauthoryear{Collier Cameron et 
al.}{2010}]{cameron2010rm} Collier Cameron A., Bruce V.~A., Miller 
G.~R.~M., Triaud A.~H.~M.~J., Queloz D., 2010, MNRAS, 403, 151

\bibitem[\protect\citeauthoryear{Demarque et 
al.}{2004}]{demarque2004} Demarque P., Woo J.-H., Kim Y.-C., Yi 
S.~K., 2004, ApJS, 155, 667 

\bibitem[\protect\citeauthoryear{Deming et al.}{2005}]{deming2005} 
Deming D., Seager S., Richardson L.~J., Harrington J., 2005, Natur, 434, 
740
 

\bibitem[\protect\citeauthoryear{Donati et al.}{1997}]{donati97zdi} 
Donati J.-F., Semel M., Carter B.~D., Rees D.~E., Collier Cameron A., 1997, 
MNRAS, 291, 658 

\bibitem[\protect\citeauthoryear{Eggleton 
\& Kiseleva-Eggleton}{2001}]{eggleton2001} Eggleton P.~P., Kiseleva-Eggleton L., 2001, ApJ, 562, 1012 



\bibitem[\protect\citeauthoryear{Enoch et al.}{2010}]{enoch2009}
Enoch, B., Collier Cameron, A., Hebb, L., Parley, N. R., 2010, A\&A, submitted

\bibitem[\protect\citeauthoryear{Fabrycky 
\& Tremaine}{2007}]{fabrycky2007kozai} Fabrycky D., Tremaine S., 2007, ApJ, 669, 1298 

\bibitem[\protect\citeauthoryear{Fekel, Warner, 
\& Kaye}{2003}]{fekel2003} Fekel F.~C., Warner P.~B., Kaye A.~B., 2003, AJ, 125, 2196 

\bibitem[\protect\citeauthoryear{Fortney, Marley, 
\& Barnes}{2007}]{fortney2007radii} Fortney J.~J., Marley M.~S., Barnes J.~W., 2007, ApJ, 659, 1661 


\bibitem[\protect\citeauthoryear{Fossey, Waldmann, 
\& Kipping}{2009}]{fossey2009hd80606} Fossey S.~J., Waldmann I.~P., Kipping D.~M., 2009, MNRAS, 396, L16 

\bibitem[\protect\citeauthoryear{Fressin et 
al.}{2007}]{fressin2007radii} Fressin F., Guillot T., Morello V., Pont F., 2007, A\&A, 475, 729 


\bibitem[\protect\citeauthoryear{{Girardi}, {Bressan}, {Bertelli} \&
  {Chiosi}}{{Girardi} et~al.}{2000}]{girardi2000}
{Girardi} L.,  {Bressan} A.,  {Bertelli} G.,    {Chiosi} C.,  2000, A\&AS, 141,
  371

\bibitem[\protect\citeauthoryear{Grenier et 
al.}{1999}]{grenier99} Grenier S., et al., 1999, A\&AS, 137, 451 

\bibitem[\protect\citeauthoryear{Guenther et 
al.}{2007}]{guenther2007} Guenther E.~W., Esposito M., Mundt R., Covino E., Alcal{\'a} J.~M., Cusano F., Stecklum B., 2007, A\&A, 467, 1147
 

\bibitem[\protect\citeauthoryear{Guenther et 
al.}{2009}]{guenther2009} Guenther E.~W., Hartmann M., Esposito M., Hatzes A.~P., Cusano F., Gandolfi D., 2009, A\&A, 507, 1659 


\bibitem[\protect\citeauthoryear{Guillot 
\& Showman}{2002}]{guillot2002irrad} Guillot T., Showman A.~P., 2002, A\&A, 385, 156 

\bibitem[\protect\citeauthoryear{Hatzes et 
al.}{2005}]{hatzes2005giant} Hatzes A.~P., Guenther E.~W., Endl M., Cochran W.~D., D{\"o}llinger M.~P., Bedalov A., 2005, A\&A, 437, 743 

\bibitem[\protect\citeauthoryear{Hatzes 
\& Zechmeister}{2007}]{hatzes2007} Hatzes A.~P., Zechmeister M., 2007, ApJ, 670, L37 


\bibitem[\protect\citeauthoryear{Hauck 
\& Mermilliod}{1997}]{hauck97} Hauck B., Mermilliod M., 1997, yCat, 2215, 0 


\bibitem[\protect\citeauthoryear{Hebb et al.}{2009}]{hebb2009wasp12} 
Hebb L., et al., 2009, ApJ, 693, 1920 



\bibitem[\protect\citeauthoryear{Ida 
\& Lin}{2005}]{ida2005} Ida S., Lin D.~N.~C., 2005, ApJ, 626, 1045 

\bibitem[\protect\citeauthoryear{Jenkins et 
al.}{2010}]{jenkins2010} Jenkins J.~M., et al., 2010, arXiv, 
arXiv:1001.0416 


\bibitem[\protect\citeauthoryear{Johnson et 
al.}{2007}]{johnson2007giants} Johnson J.~A., et al., 2007, ApJ, 665, 785 

\bibitem[\protect\citeauthoryear{Kane et al.}{2008}]{kane2008} 
Kane S.~R., et al., 2008, MNRAS, 384, 1097 

\bibitem[\protect\citeauthoryear{Kennedy 
\& Kenyon}{2008}]{kennedy2008snowline} Kennedy G.~M., Kenyon S.~J., 2008, ApJ, 673, 502 

\bibitem[\protect\citeauthoryear{Kozai}{1962}]{kozai62} Kozai 
Y., 1962, AJ, 67, 591 


\bibitem[\protect\citeauthoryear{Knutson et 
al.}{2007}]{knutson2007} Knutson H.~A., et al., 2007, Natur, 447, 
183 


\bibitem[\protect\citeauthoryear{Kunzli et 
al.}{1997}]{kunzli97cal} Kunzli M., North P., Kurucz R.~L., Nicolet B., 1997, A\&AS, 122, 51 

\bibitem[\protect\citeauthoryear{Lin, Bodenheimer, 
\& Richardson}{1996}]{lin96} Lin D.~N.~C., Bodenheimer P., Richardson D.~C., 1996, Natur, 380, 606 

\bibitem[\protect\citeauthoryear{Lister et al.}{2007}]{lister2007} 
Lister T.~A., et al., 2007, MNRAS, 379, 647 

\bibitem[\protect\citeauthoryear{Mandel 
\& Agol}{2002}]{mandel2002} Mandel K., Agol E., 2002, ApJ, 580, L171 

\bibitem[\protect\citeauthoryear{Mandushev et 
al.}{2007}]{mandushev2007tres4} Mandushev G., et al., 2007, ApJ, 667, L195 



\bibitem[\protect\citeauthoryear{Murdoch, Hearnshaw, 
\& Clark}{1993}]{murdoch93} Murdoch K.~A., Hearnshaw J.~B., Clark M., 1993, ApJ, 413, 349 


\bibitem[\protect\citeauthoryear{Nagasawa, Ida, 
\& Bessho}{2008}]{nagasawa2008kozai} Nagasawa M., Ida S., Bessho T., 2008, ApJ, 678, 498 

\bibitem[\protect\citeauthoryear{Narita et al.}{2009}]{narita2009_hat-p-7} 
Narita N., Sato B., Hirano T., Tamura M., 2009, PASJ, 61, L35 

\bibitem[\protect\citeauthoryear{Ogilvie 
\& Lin}{2007}]{ogilvie207tide} Ogilvie G.~I., Lin D.~N.~C., 2007, ApJ, 661, 1180 

\bibitem[\protect\citeauthoryear{Perryman et 
al.}{1997}]{perryman97} Perryman M.~A.~C., et al., 1997, A\&A, 323, L49 




\bibitem[\protect\citeauthoryear{Pollacco et 
al.}{2006}]{pollacco2006} Pollacco D.~L., et al., 2006, PASP, 118, 
1407 

\bibitem[\protect\citeauthoryear{Pollacco et 
al.}{2008}]{pollacco2007_wasp-3} Pollacco D., et al., 2008, MNRAS, 385, 
1576 

\bibitem[\protect\citeauthoryear{Rasio 
\& Ford}{1996}]{rasio96} Rasio F.~A., Ford E.~B., 1996, Sci, 274, 954 

\bibitem[\protect\citeauthoryear{Sato et al.}{2008}]{sato2008giants} 
Sato B., et al., 2008, PASJ, 60, 1317 


\bibitem[\protect\citeauthoryear{Seager 
\& Mall{\'e}n-Ornelas}{2003}]{seager2003} Seager S., Mall{\'e}n-Ornelas G., 2003, ApJ, 585, 1038 

\bibitem[\protect\citeauthoryear{Smalley, Smith, 
\& Dworetsky}{2001}]{smalley2001uclsyn} Smalley B., Smith K. C., Dworetsky M. M., 2001, UCLSYN Userguide

\bibitem[\protect\citeauthoryear{Smalley}{1993}]{smalley93} Smalley B., 1993, A\&A, 274, 391 

\bibitem[\protect\citeauthoryear{Smith}{1992}]{smith92uclsyn} Smith K.C., 1992, PhD thesis, University College, London

\bibitem[\protect\citeauthoryear{Snellen et 
al.}{2009}]{snellen2009} Snellen I.~A.~G., et al., 2009, A\&A, 497, 545 

\bibitem[\protect\citeauthoryear{Sozzetti et 
al.}{2007}]{sozzetti2007} Sozzetti A., Torres G., Charbonneau D., 
Latham D.~W., Holman M.~J., Winn J.~N., Laird J.~B., O'Donovan F.~T., 2007, 
ApJ, 664, 1190 

\bibitem[\protect\citeauthoryear{Street et al.}{2007}]{street2007} 
Street R.~A., et al., 2007, MNRAS, 379, 816 


\bibitem[\protect\citeauthoryear{Torres, Andersen, 
\& Gim{\'e}nez}{2009}]{torres2009} Torres G., Andersen J., Gim{\'e}nez A., 2009, A\&ARv, 13 

\bibitem[\protect\citeauthoryear{Uytterhoeven et 
al.}{2008}]{uytterhoeven2008puls} Uytterhoeven K., et al., 2008, A\&A, 489, 1213 

\bibitem[\protect\citeauthoryear{Vogt 
\& Penrod}{1983}]{vogt83doppler} Vogt S.~S., Penrod G.~D., 1983, PASP, 95, 565 

\bibitem[\protect\citeauthoryear{Weidenschilling 
\& Marzari}{1996}]{weidenschilling96} Weidenschilling S.~J., Marzari F., 1996, Natur, 384, 619 


\bibitem[\protect\citeauthoryear{Willems 
\& Aerts}{2002}]{willems2002} Willems B., Aerts C., 2002, A\&A, 384, 441 

\bibitem[\protect\citeauthoryear{Winn et al.}{2009}]{winn2009_hat-p-7} 
Winn J.~N., Johnson J.~A., Albrecht S., Howard A.~W., Marcy G.~W., 
Crossfield I.~J., Holman M.~J., 2009, ApJ, 703, L99 

\bibitem[\protect\citeauthoryear{Wolff}{1983}]{wolff83} Wolff 
S.~C., 1983, NASSP, 463,  


\bibitem[\protect\citeauthoryear{Wu 
\& Murray}{2003}]{wu2003kozai} Wu Y., Murray N., 2003, ApJ, 589, 605 


\end{thebibliography}

\bsp

\label{lastpage}

\end{document}